\documentclass[aps, prb, superscriptaddress, reprint, nobibnotes, twocolumn, longbibliography]{revtex4-2}
\usepackage{textcomp}
\usepackage{amsmath,amsthm,amssymb,mathtools,physics} 
\usepackage{xcolor} 
\usepackage{graphicx} 
\usepackage[font=small, justification=raggedright, singlelinecheck=false]{caption}
\usepackage[utf8]{inputenc} 

\usepackage{subcaption} 
\usepackage{bm} 
\usepackage{lineno} 
\usepackage{geometry} 
\usepackage{adjustbox} 
\usepackage{placeins} 
\usepackage[T1]{fontenc} 
\usepackage{lipsum} 
\usepackage{csquotes} 
\usepackage{enumerate} 
\usepackage{epstopdf} 
\usepackage{fancyhdr} 
\usepackage{marvosym} 
\usepackage{wrapfig} 
\usepackage{indentfirst} 
\usepackage{balance} 

\def\bef{\begin{framed}}
\def\eef{\end{framed}}
\def\be{\begin{equation}}
\def\ee{\end{equation}}
\def\ber{\begin{eqnarray}}
\def\eer{\end{eqnarray}}

\usepackage[pdftex, pdftitle={Article}, pdfauthor={Author}]{hyperref}
\hypersetup{breaklinks=true, citecolor=blue, colorlinks=true, linkcolor=red}

\begin{document}
\title{Low-Field Metal-Insulator Transition in AB-Stacked Bilayer Graphene}

\author{Amarnath Chakraborty}
\email{amarnathchakraborty95@gmail.com}
\affiliation{Department of Physics and Astronomy, University of Missouri,
             Columbia, Missouri, USA}
\author{Aleksandr Rodin}
\affiliation{Yale-NUS College, 16 College Avenue West, 138527, Singapore}
\affiliation{Centre for Advanced 2D Materials, National University of
             Singapore, 117546, Singapore}
\affiliation{Department of Materials Science and Engineering,
             National University of Singapore, 9 Engineering Drive 1,
             117575, Singapore}
\author{Shaffique Adam}
\affiliation{Department of Physics, Washington University in St.\ Louis,
             St.\ Louis, Missouri 63130, United States}
\author{Giovanni Vignale}
\email{vgnl.g@nus.edu.sg}
\affiliation{The Institute for Functional Intelligent Materials (I-FIM),
             National University of Singapore, 4 Science Drive 2,
             Singapore 117544}
\date{\today}

\begin{abstract}
We investigate the interplay of in-plane magnetic and transverse
electric fields in AB-stacked bilayer graphene. In prior work neglecting
trigonal warping, we demonstrated that this configuration induces an
insulator-metal transition purely via orbital effects, albeit requiring
impractically large magnetic fields ($>100$\,T). Here, we extend the
analysis to the ultra-low-energy regime by incorporating interlayer skew
couplings. In a restricted region of momentum space, trigonal warping
produces a fine splitting of Dirac cones leading to a compensated
semimetallic state at zero external field. Application of a transverse
electric field above a small threshold ($V_c\sim0.6$\,meV) reinstates an
insulating gap. When an in-plane
magnetic field is applied, the orbital gauge vector immediately breaks the
$C_3$ spatial symmetry of the lattice, and we uncover two sequential,
field-driven transitions separated by an order of magnitude in scale.
First, at fields ($B \approx 1$\,T), the Zeeman effect
drives an indirect insulator-to-semimetal transition at the ungated
charge neutrality point; the cross-spin gap closes, generating distinct
electron and hole pockets separated in momentum space. Second, as the
field increases to $B \approx 10$--$25$\,T, the orbital coupling closes
the same-spin gap at an energy well away from the charge-neutral Fermi
level. By electrostatically tuning the Fermi level to this specific
gap-closing energy, we reveal the emergence of $C_3$-broken same-spin
Fermi pockets, accompanied by a distinct step-like onset in the density
of states. This dual-transition regime provides a highly sensitive
platform for tunable, spin-selective transport.
\end{abstract}

\maketitle

\section{Introduction}

The electronic and magnetic properties of graphene systems under magnetic
fields have long been a topic of great interest in solid state physics.
In monolayer graphene, perpendicular magnetic fields produce striking
phenomena such as relativistic Landau levels~\citep{Zhang2005}, integer
\citep{Novoselov2006} and fractional~\citep{Bolotin2009} quantum Hall
effects, and gate-tunable orbital magnetism.

Bilayer graphene (BLG) offers an even richer scenario.
Interlayer coupling and stacking geometry enable mechanical, electrical,
and magnetic control over band topology. For instance, twisted bilayer
graphene hosts flat bands that give rise to superconductivity and
correlated insulating states~\citep{Cao2018}, while Bernal-stacked
bilayers exhibit gate-tunable bandgaps~\citep{Seiler2024} and orbital
magnetism sensitive to in-plane magnetic fields ($B_\parallel$)
\citep{Kammermeier_2019,PhysRevB.89.125418,PhysRevB.111.125130}.
The latter arises because $B_\parallel$ couples asymmetrically to the
two layers, shifting their uncoupled band structures in opposite momentum
directions. 

\begin{figure*}
\centering
\includegraphics[width=\textwidth]{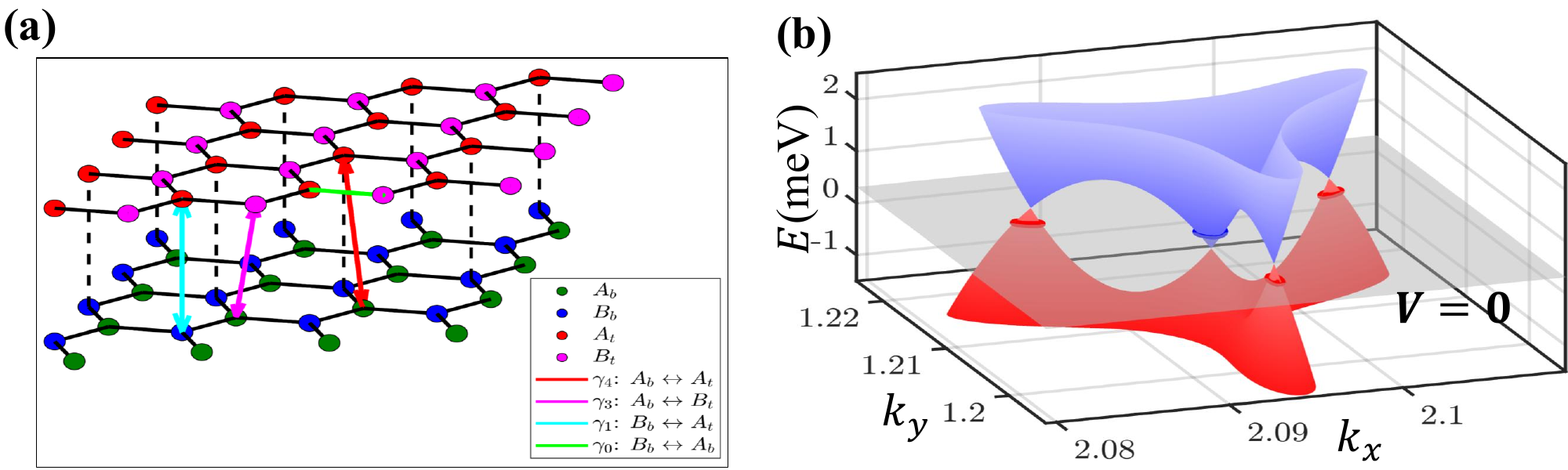}
\caption{AB-stacked bilayer graphene lattice and band structure with
trigonal warping. \textbf{(a)}~Side view of the AB-stacked lattice
showing relevant interlayer hopping terms. \textbf{(b)}~Semi-metallic
state at zero field with $E_F=0.313\,\mathrm{meV}$. Note that the Fermi level is shifted from zero energy due to particle-hole asymmetry.}
\label{fig:Bands}
\end{figure*}

On the other hand, the low-energy distortion of the Fermi surface
induced by interlayer skew couplings ($\gamma_3$, $\gamma_4$) in the
Slonczewski--Weiss--McClure
model~\citep{PhysRev.109.272,PhysRev.119.606} renormalizes the Fermi
velocity and imparts a threefold symmetry to the low-energy band
structure~\citep{McCann_2013,Mucha-Kruczyński_2010,Seiler2024}.
Tight-binding models that include skew hoppings ($\gamma_3,\gamma_4$)
predict a gapless fine structure featuring four mini-Dirac cones and
van Hove singularities at modest displacement
fields~\citep{Weitz_2010,Novoselov2005,partoens2006graphene}.
Furthermore, this warped low-energy Fermi surface yields anomalous
quantum-Hall sequences in
BLG~\citep{PhysRevLett.96.086805,Seiler2022,zhou2022isospin,varlet2015tunable},
as well as in Bernal trilayer
graphene~\citep{zibrov2018emergent,winterer2022spontaneous,campos2012quantum},
rhombohedral trilayer
graphene~\citep{zhou2021half,zhou2021superconductivity}, and Bernal
tetralayer graphene~\citep{shi2018tunable}.

Recently, Seiler {\it et al.}~\citep{Seiler2024} investigated the fine
structure of trigonal warping under an applied displacement field, both
theoretically and experimentally, addressing the long-standing absence
of clear signatures of linearly dispersing bands and
electric-field-driven topological transitions in BLG. By combining
high-field magnetotransport measurements on hBN-encapsulated bilayer
graphene with realistic tight-binding modeling~\cite{PhysRevB.89.035405},
they identified signatures of band-structure reconstruction associated
with a van Hove singularity. In particular, the emergence of four
Dirac cones leads to an unconventional sequence of Landau levels,
distinct from the previously established picture.

\begin{figure*}
\centering
\includegraphics[width=\textwidth]{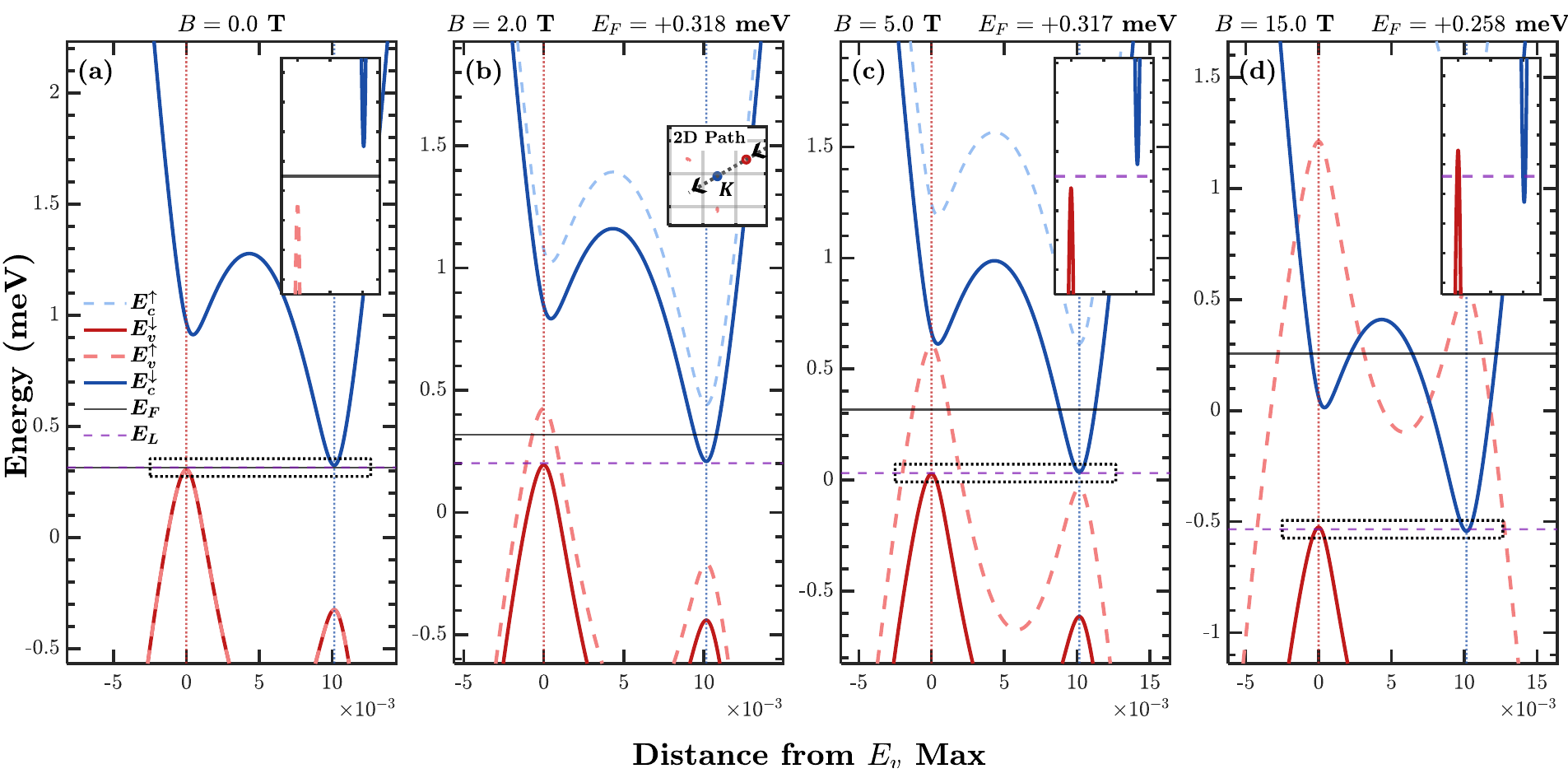}
\caption{Evolution of the low-energy 1D band structure and topological
transitions under an in-plane magnetic field at $V=0.65\,\mathrm{meV}$.
\textbf{(a)}~$B=0\,\mathrm{T}$: The initial insulating state. The
ungated Fermi level ($E_F$) sits within an open cross-spin gap, as
magnified in the top-right inset.
\textbf{(b)}~$B=2\,\mathrm{T}$: The Zeeman-driven spin-polarized
semimetal. The cross-spin gap has closed at $E_F$. The dotted line in the inset
illustrates the trajectory along which the bands are plotted running from the red dot (red vertical line) to the blue dot (blue vertical line). The central blue dot marks the K point and the red  markers indicate the positions of the hole pockets shown in Fig. 1(b). 
\textbf{(c)}~$B=5\,\mathrm{T}$: Intermediate orbital distortion. While
the system is metallic at $E_F$, the same-spin gap remains open at the
critical crossing energy ($E_L$), highlighted in the inset.
\textbf{(d)}~$B=15\,\mathrm{T}$: The orbital-driven semimetallic phase.
The orbital Peierls phase forces the same-spin gap to close exactly at
$E_L$ (inset), marking a symmetry-breaking topological transition that
can be accessed by gating the system to this specific energy.}
\label{fig:1DBands}
\end{figure*}

Despite these advances, the combined effects of an in-plane magnetic
field ($B_\parallel$) and a transverse electric field ($E_\perp$) in
the presence of trigonal warping remain largely unexplored.
In our prior work~\citep{PhysRevB.111.125130}, we showed that
$B_\parallel$ and $E_\perp$ together can drive an insulator-metal (IM)
transition purely via orbital effects, albeit requiring
impractically large magnetic fields ($>100$\,T). Here, we extend that
study by incorporating skew hoppings ($\gamma_3$ and $\gamma_4$) into full
tight-binding simulations of the $B_\parallel$--$E_\perp$ transition
(Fig.~\ref{fig:Bands}a--b). Focusing on the low-energy Fermi surface within
roughly $1\%$ of the Brillouin zone (within $\sim1$\,meV of the Dirac points),
we confirm that the system evolves from a gapless semimetal to a gapped
insulator above a critical displacement field, consistent with Ref.~\citep{Seiler2024}.
Crucially, we find that this warped geometry radically alters the magnetic
response: it not only brings the critical magnetic field down to an experimentally
accessible regime ($<30$\,T)~\citep{doi:10.1126/science.abm8386}, but splits
the IM phase boundary into a striking dual-transition sequence.

This sequential transition arises from a fundamental inversion of energy
scales. In our earlier, unwarped study, the native gap (on the order of
hundreds of meV) so far exceeded the Zeeman energy ($2\mu_B B \sim 0.1\,\mathrm{meV/T}$)
that spin entered only as a minor correction. Trigonal warping changes
this hierarchy entirely: by reducing the native gap to a fraction of a meV,
it elevates the Zeeman effect to a primary driver. Consequently, the
initial insulator-to-metal transition occurs at a rather low magnetic field
($B_c^Z\sim1\,\mathrm{T}$), where Zeeman splitting closes the cross-spin
gap. This converts the insulator into a spin-polarized semimetal precisely
at the ungated charge neutrality point, generating a Fermi surface of
distinct electron and hole pockets.

An order of magnitude higher in field ($B_c^{\rm orb}\sim10\text{--}25\,\mathrm{T}$),
the orbital vector potential takes over, heavily distorting the band
structure to drive a subsequent semimetallic transition via the closure of
the \emph{same-spin} gap. The defining physical consequence of this
asymmetric orbital coupling is that the same-spin gap closes roughly
$1\,\mathrm{meV}$ \emph{away} from the charge-neutral Fermi level. Because
this second topological event occurs entirely outside the ungated transport
window, observing the resulting same-spin pocket reconstruction requires explicitly
gating the system. By electrostatically tuning the Fermi level to match the
exact energy where the same-spin bands overlap, we reveal a highly sensitive,
field- and gate-tunable route to spin-selective transport signatures.

\section{Tight Binding Model}

The Bernal-stacked graphene bilayer is described by a tight-binding
Hamiltonian with nearest-neighbor intra- and interlayer
hopping~\cite{RevModPhys.81.109}:
\begin{align}
H_{\mathbf{k}} =
\begin{pmatrix}
  -V/2 & \gamma_{0}f(\mathbf{k}_\mathrm{t}) &
  \gamma_4 f(\mathbf{k}) & \gamma_3 f^*(\mathbf{k}) \\
  \gamma_{0}f^\dagger(\mathbf{k}_\mathrm{t}) & -V/2+\Delta &
  \gamma_{1} & \gamma_4 f(\mathbf{k}) \\
  \gamma_4 f^*(\mathbf{k}) & \gamma_{1} & V/2+\Delta &
  \gamma_{0}f(\mathbf{k}_\mathrm{b}) \\
  \gamma_3 f(\mathbf{k}) & \gamma_4 f^*(\mathbf{k}) &
  \gamma_{0}f^\dagger(\mathbf{k}_\mathrm{b}) & V/2
\end{pmatrix}
\label{eqn:FullHam}
\end{align}
The four basis states are $\mathrm{A}_\mathrm{t}$, $\mathrm{B}_\mathrm{t}$,
$\mathrm{A}_\mathrm{b}$, and $\mathrm{B}_\mathrm{b}$ with $\mathrm{A}$ and
$\mathrm{B}$ denoting the sublattice and $\mathrm{t}/\mathrm{b}$
subscripts identifying the top and bottom layers.
The hopping energy between $\mathrm{B}_\mathrm{t}$ and $\mathrm{A}_\mathrm{b}$
is $\gamma_1=0.361\,\mathrm{eV}$, while the nearest-neighbour intralayer
hopping is $\gamma_0=-2.61\,\mathrm{eV}$~\cite{PhysRevB.89.035405,Seiler2024}.
Here $f(\mathbf{k})=\sum_{n=1}^3 e^{i\mathbf{k}\cdot\boldsymbol{\delta}_n}$,
where $\boldsymbol{\delta}_1=a(-1,\sqrt{3})/2$,
$\boldsymbol{\delta}_2=a(-1,-\sqrt{3})/2$, and
$\boldsymbol{\delta}_3=a(1,0)$ are the intra-layer nearest-neighbour
vectors and $a\approx1.42$\,\AA\ is the carbon bond length.
The diagonal terms $\pm V$ originate from an external electric field
applied perpendicular to the bilayer.
We also include skew interlayer couplings $\gamma_3=-0.283\,\mathrm{eV}$,
$\gamma_4=0.138\,\mathrm{eV}$, and the dimer/non-dimer energy offset
$\Delta=0.015\,\mathrm{eV}$~\cite{PhysRevB.89.035405,Seiler2024}.

The in-plane magnetic field $\mathbf{B}_\parallel$ introduces two
distinct physical effects: an orbital Peierls phase and a Zeeman spin
splitting. The orbital vector potential enters via the Peierls
substitution $\mathbf{k}\to\mathbf{k}\pm e\mathbf{A}/\hbar$, giving
\begin{equation}
  \mathbf{k}_\mathrm{t/b}\,a
  = \mathbf{k}\,a \pm \pi\Phi(\sin\theta,-\cos\theta,0),
  \label{eqn:Peierls}
\end{equation}
where $\ell$ is the interlayer separation and
$\Phi=Ba\ell/\Phi_0$ is the magnetic flux in units of
$\Phi_0=h/e\simeq4.14\times10^5\,\mathrm{T\,}\text{\AA}^2$.
Here $\theta=\pi/2$ corresponds to a field along the armchair direction.
The skew-coupling terms $\gamma_3,\gamma_4$ carry no Peierls
phase: in the gauge $\mathbf{A}=B_y z\,\hat{x}$, the phase accumulated
along an interlayer hop from $z=+\ell/2$ to $z=-\ell/2$ vanishes
exactly, because the gauge field is odd under $z\to-z$ and the hop path
is symmetric about $z=0$; a full line-integral evaluation is given in
the SM~\citep{SI}.

Because the in-plane field also couples directly to the electron spin,
we explicitly define the full crossed-field Hamiltonian as:
\begin{equation}
  H^\sigma_\mathbf{k} = H_\mathbf{k} + \sigma\mu_B B\,\mathbf{I}_{4\times4},
  \label{eqn:Zeeman}
\end{equation}
where $\sigma=+1$ for spin-$\uparrow$, $\sigma=-1$ for spin-$\downarrow$,
and $\mu_B$ is the Bohr magneton. Unlike the orbital term, the Zeeman
term acts as a uniform energy shift independent of $\mathbf{k}$.
We show in Sec.~\ref{sec:low_energy} that these two field-coupling mechanisms operate at drastically different energy scales and this separation of scales is the direct origin of the two distinct critical fields
($B_c^Z\ll B_c^{\rm orb}$) that govern the topological transitions in
this system.

\section{Accessible IM Transition Enabled by Trigonal Warping}
\label{sec:low_energy}

In our previous work~\citep{PhysRevB.111.125130}, we analyzed the
insulator-semimetal transition in AB-stacked bilayer graphene at charge neutrality under a transverse field $E_\perp$ and an in-plane magnetic
flux $\Phi$. By including only the interlayer coupling $\gamma_1$, and neglecting the Zeeman coupling, 
we found that the system remains gapped as long as
$3\pi|\gamma_0|\Phi<\gamma_1$
and transitions to a compensated semimetal when $\Phi$ exceeds
\begin{equation}
  \Phi_c = \frac{\gamma_1}{3\pi|\gamma_0|},
\end{equation}
corresponding to $B_c\approx100\,\mathrm{T}$ (with auxiliary strain).
In the present work, we show that the trigonal-warping terms dramatically
change this scenario.

\subsection{Critical displacement field and gap opening}

Following Ref.~\citep{McCann_2013} one can show that at low energy the
trigonal warping fragments the Fermi surface into one central pocket
(conduction band) and three satellite pockets (valence band) around each
$K$ point (Fig.~\ref{fig:Bands}(b)). By resolving the band structure on
a scale of $1/100^\text{th}$ of the original Brillouin zone (BZ), one
uncovers threefold-symmetric Dirac cones which cross the Fermi level,
giving rise to a semimetallic state at $V=0$. Due to the particle-hole asymmetry introduced by $\gamma_4$, a perpendicular displacement field $E_\perp$ does not shift the central Dirac cone and the satellite pockets symmetrically. Because these extrema evolve at different rates, the system remains in a compensated semimetallic state at low fields. A true bulk insulating gap is observed only once the displacement field exceeds a critical threshold,
\begin{equation}
  V_c \approx 0.63\,\mathrm{meV}\,.
  \label{eqn:D_c}
\end{equation}
We obtain the value of $V_c$ by direct diagonalization of the full tight-binding
Hamiltonian (Eq.~\ref{eqn:FullHam}) on a dense $k$-grid; the numerical
procedure is detailed in the Supplemental Material~\footnote{See
\cite{SM} for the numerical procedure for phase boundary determination,
the Zeeman analysis, the angular dependence of $B_c$, and parameter
robustness checks.}.

This fundamentally alters the scenario of the magnetic field induced
insulator-metal (IM) transition. Below $V_c$, residual valence-band
pockets sustain a finite density of states at the Fermi level (shifted
to $E_F\approx0.31\,\mathrm{meV}$ by particle-hole asymmetry from
$\gamma_4$), whereas for $V>V_c$ all pockets are gapped and a full bulk
gap appears. The small insulating gap can be closed by a relatively
small in-plane magnetic field ($\approx10\,\mathrm{T}$), leading to an
IM transition, as we discuss in the following section.

Because $V_c$ depends sensitively on the skew-hopping ratio
$\gamma_3/\gamma_1$, our results suggest that strain or dielectric
engineering could tune the threshold, offering an additional control
parameter for low-field IM transitions.

This behavior has two key implications. First, it highlights the
necessity of including skew-hopping terms ($\gamma_3,\gamma_4$) when
modeling low-energy transport and optics: below $V_c$, the ungapped
pockets dominate conductivity and yield anisotropic cyclotron orbits in
quantum-oscillation measurements. Second, the Berry-curvature textures
of these warped pockets---even prior to full gap opening---may drive
pronounced valley-Hall or nonlinear Hall responses that switch sharply
at $V_c$.

With $V_c$ established as the zero-field gap threshold, we can now
explore how an in-plane magnetic flux $\Phi$ further suppresses or
closes the gap, thereby tracing out the IM phase boundary in the
$(V,\Phi)$ plane.

\subsection{Phase boundary and gap scaling}

\begin{figure*}[ht]
  \centering
  \includegraphics[width=\textwidth]{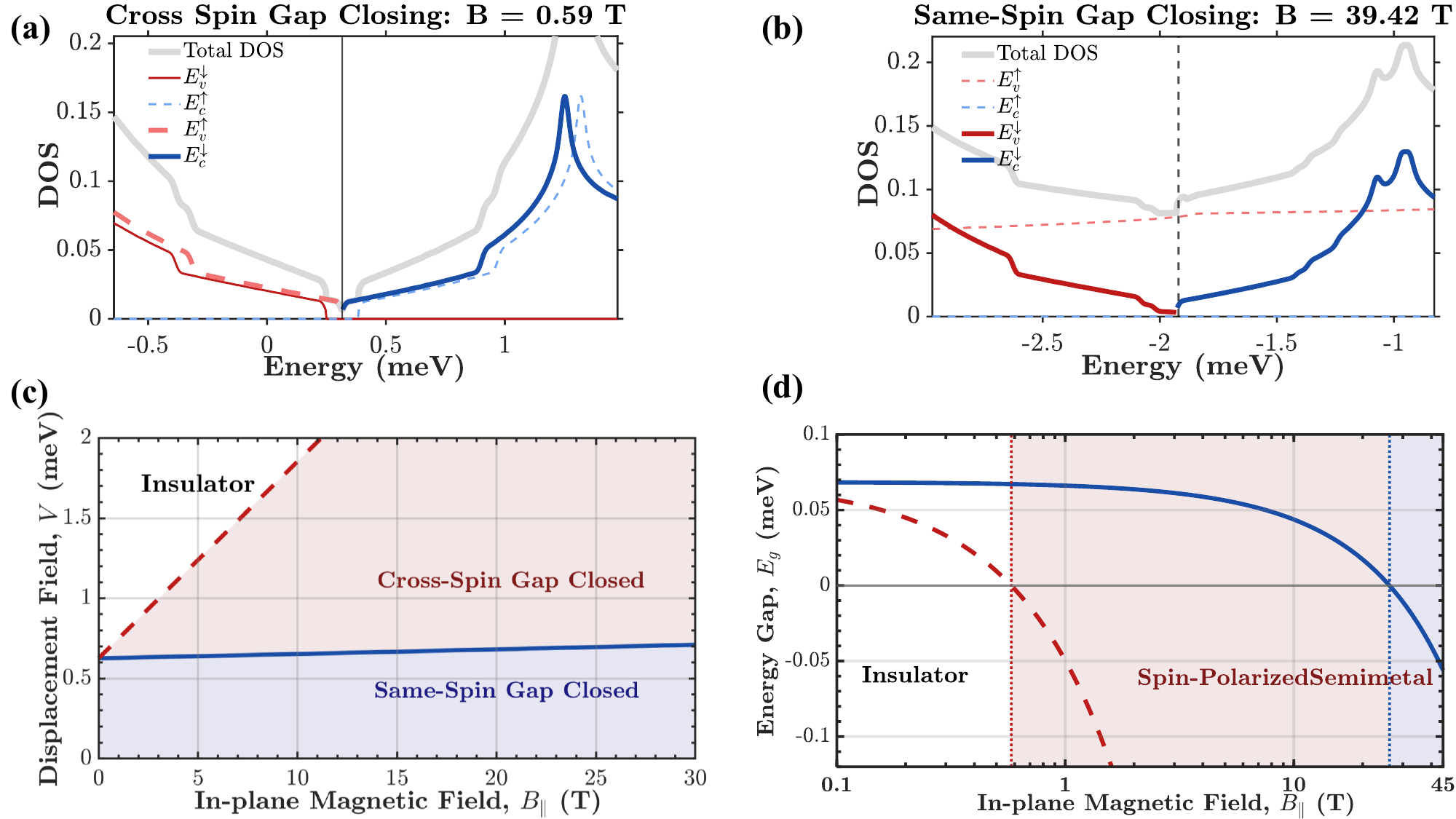}
  \caption{
    \textbf{(a)}~Density of states (DOS) across the Zeeman transition ($B = 0.58\,\mathrm{T}$, $V = 0.7\,\mathrm{meV}$). The cross-spin bands ($E_v^\uparrow$, dashed red; $E_c^\downarrow$, dotted blue) close the gap precisely at the ungated Fermi level $E_F$ (solid vertical line), driving the transition to a spin-polarized semimetal. 
    \textbf{(b)}~DOS in the orbital-driven semimetallic regime ($B = 39.36\,\mathrm{T}$, $V = 0.7\,\mathrm{meV}$). The spin-down bands indicated by red solid line (valence) and blue solid line (conduction)  come together at the critical crossing energy $E_L$ (dashed vertical line) producing a small step-like feature in the density of states. This topological feature forms well below $E_F$ and can only be accessed by electrostatic gate tuning.
    \textbf{(c)}~$V$--$B_\parallel$ phase diagram outlining the cross-spin (dashed red) and same-spin (solid blue) gap closures. The dashed red line separates the insulator (white) from the spin-polarized semimetal.  The semimetallic region is further split into a top region (red shading), in which only the cross-spin gap is closed and a lower region (blue shading) in which the same-spin gap is also closed.
    \textbf{(d)}~Cross-spin gap (dashed red line) and same-spin gap (blue solid line) versus $B_\parallel$ (log scale) at $V = 0.7\,\mathrm{meV}$. The white region of the graph, for $B< 0.6\,\mathrm{T}$, is where both gaps are open causing the system to be an insulator; in the red-shaded region $0.6\,\mathrm{T}<B <26.2\,\mathrm{T}$ the cross-spin gap is closed due to the Zeeman effect, causing the system to become a semimetal;  in the blue-shaded region, $B> 26.2\,\mathrm{T}$, the  same-spin gap is also closed due to the orbital effect.}
  \label{fig:V_vs_Phi}
\end{figure*}

To investigate the metal-insulator transition we first examine the
density of states (DoS) under the two relevant field configurations,
as shown in Fig.~\ref{fig:V_vs_Phi}(a,b) --- discussed in detail in
Sec.~\ref{sec:zeeman_first} and Sec.~\ref{sec:orbital_second} below. To
capture the complete transition, we also construct a $V$--$B$ phase
diagram (Fig.~\ref{fig:V_vs_Phi}(c)). The full phase diagram extending
to higher fields and connecting to the unwarped high-field regime of
Ref.~\citep{PhysRevB.111.125130} is shown in Fig.~S1 of the~\cite{SM}.

With the phase boundary determined, we convert the critical flux into a
physical field via $B_c(V)=\Phi_c(V)/(\Phi_0a\ell)$. 
When trigonal warping is included, the orbital gap-closing field falls to the $\mathcal{O}(10\,\mathrm{T})$ range---roughly an order of magnitude smaller than the unwarped estimate ($B_c\gtrsim100\,\mathrm{T}$). Furthermore, the reduced native gap elevates the Zeeman effect to a primary driver of the phase diagram, initiating a separate spin-polarized transition at even lower fields. This
dramatic reduction has a numerical signature through an abrupt jump of
the density of states at the Fermi level once $\Phi>\Phi_c(V)$,
signalling the sudden formation of compensated electron-hole pockets.

Quantitatively, in the low-energy regime a displacement field $V>V_c$
opens an indirect gap $\Delta(V)\propto V-V_c$, defined as the
difference between the minimum of the conduction band and the maximum
of the valence band. This behavior differs qualitatively from our
earlier high-energy analysis, where the zero-field indirect gap was
essentially set by the interlayer coupling $\gamma_1$ and insensitive
to $V$. In the present, warped low-energy band structure the zero-field
gap is much smaller (fractions of a meV rather than hundreds
of meV) and scales with $V-V_c$ rather than being fixed by $\gamma_1$.

It is useful to parametrize the field dependence of the gap as
\begin{equation}
  E_g(B;\,V) = \Delta(V) - \beta(V)\,B,
  \label{eq:EgB}
\end{equation}
where $\beta(V)$ is a $V$-dependent gap-closing rate (units: meV/T). It can be
extracted from the blue curve (Orbital gap closing) of Fig.~\ref{fig:V_vs_Phi}(d). At
$V=0.7\,\mathrm{meV}$ we obtain
$\beta\approx3.1\times10^{-3}\,\mathrm{meV/T}$. Compared to orbital gap-closing rate the Zeeman scale is $2\mu_B=0.1158\,\mathrm{meV/T}$ (extracted from the red curve)---larger
by a factor of roughly $38$.

A low-energy two-band
analysis~\cite{SM} yields the leading-order analytical expression
\begin{equation}
  \beta \approx \frac{4Vv_3}{\gamma_1}\,\frac{\pi a\ell}{\Phi_0},
  \label{eq:beta_analytic}
\end{equation}
demonstrating that $\beta\propto V$. Consequently,
\begin{equation}
  B_c \;\propto\; \frac{\Delta(V)}{\beta} \;\propto\; (V-V_c),
  \label{eq:Bc_linear}
\end{equation}
consistent with the numerical results shown in Fig.~\ref{fig:V_vs_Phi}(c); that the required magnetic field for the orbital gap closing varies almost linearly with the applied displacement field.

The low-energy IM transition persists up to a magnetic field on the
order of $10^2\,\mathrm{T}$ ($\approx137.5\,\mathrm{T}$ with our
parameter set); beyond this scale $k_B$ exceeds the characteristic
trigonal-warping momentum $k_w$, the pocket picture breaks down, and the
band structure evolves to a single anisotropic pocket. The phase
boundary (Fig.~S1 of the~\cite{SM}) then saturates toward the
constant-field behavior found in our previous high-energy
study~\citep{PhysRevB.111.125130}.

In short, including trigonal warping both (i)~reduces the zero-field
gap (via $\gamma_4$ and the pocket structure) and (ii)~converts the IM
critical field from a $V$-independent constant into an approximately
linear function of $V-V_c$.

\subsection{The Zeeman transition: insulator to spin-polarized semimetal}
\label{sec:zeeman_first}

\begin{figure*}[ht]
  \centering
  \includegraphics[width=\textwidth]{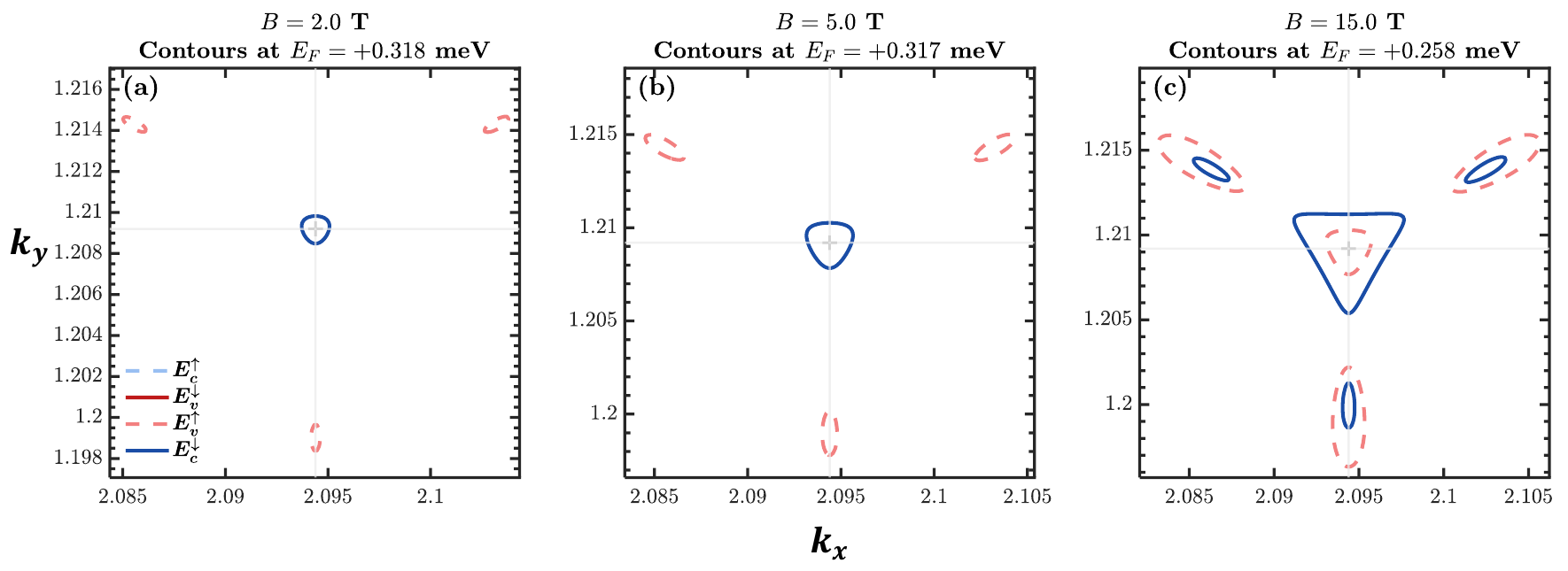}
\caption{Evolution of the 2D Fermi surface contours evaluated at the
ungated Fermi level ($E_F = +0.318\,\mathrm{meV}$) with $V=0.65\,\mathrm{meV}$. Solid dark blue
lines denote the spin-down conduction band ($E_c^\downarrow$), while dashed 
light red lines denote the spin-up valence band ($E_v^\uparrow$). The evaluation energy matches the initial state in Fig.~\ref{fig:1DBands}(a) for direct comparison.
\textbf{(a)}~$B=2.0\,\mathrm{T}$: The initial spin-polarized
semimetallic state driven by the Zeeman effect. The cross-spin gap is
closed, yielding a distinct central electron pocket ($E_c^\downarrow$) and 
three satellite hole pockets ($E_v^\uparrow$).
\textbf{(b)}~$B=5.0\,\mathrm{T}$: Intermediate regime showing pronounced 
trigonal warping and expansion of the central electron pocket, while the 
satellite hole pockets persist.
\textbf{(c)}~$B=15.0\,\mathrm{T}$: The Fermi surface at high fields. Crucially, 
the spin-down conduction band ($E_c^\downarrow$) now crosses $E_F$ at the satellite 
locations as well, spawning three new satellite electron pockets that are nested 
within the existing spin-up hole pockets. Because these contours are evaluated 
strictly at the charge-neutral $E_F$, they capture only the primary cross-spin
transport channels, while the same-spin topological Lifshitz
transition remains hidden at the lower energy ($E_L$).}
  \label{fig:FermiZeeman}
\end{figure*}

Exploiting the magnitude discrepancy established in Eq.~\ref{eqn:Zeeman}
($2\mu_B\gg\beta$), the very first field-driven transition encountered
as $B_\parallel$ increases from zero is the Zeeman crossing, occurring
at:
\begin{equation}
  B_c^Z \approx \frac{\Delta(V)}{2\mu_B} \approx 0.6\text{--}3\,\mathrm{T}
  \label{eq:BcZ}
\end{equation}
(depending on $V\in[0.7,1.0]\,\mathrm{meV}$). At this field, the
cross-spin gap $E_g(B)-2\mu_BB$ closes [Fig.~\ref{fig:V_vs_Phi}(a)],
producing a spin-polarized semimetal: only spin-$\downarrow$ electron
pockets and spin-$\uparrow$ hole pockets coexist at the Fermi
level $E_F$, while both same-spin gaps remain open (Zeeman shifts
cancel exactly in the same-spin channel; see SM Sec.~S-V.1). As the field increases, the Fermi surface undergoes a
symmetric topological evolution from one spin-$\downarrow$ conduction
and three spin-$\uparrow$ valence pockets into a four-and-four
configuration, as shown in Fig.~\ref{fig:FermiZeeman}. Because this transition occurs precisely at the ungated charge
neutrality point, it is directly observable in standard longitudinal
transport without requiring explicit spin-sensitive probes.

Because $B_c^Z\propto\Delta(V)/2\mu_B$ and the orbital correction to
the denominator is only $\beta/2\mu_B\approx2.6\%$ (see SM), $B_c^Z$ is
also far less sensitive to gate-voltage uncertainty than
$B_c^{\rm orb}$: for the same $\delta V$, the resulting uncertainty in
$B_c^Z$ is $\sim38$ times smaller.

\subsection{The orbital transition: emergence of same-spin Fermi pockets}
\label{sec:orbital_second}

An order of magnitude higher in field, at
$B_c^{\rm orb}\approx26.6\,\mathrm{T}$ (at $V=0.71\,\mathrm{meV}$), the
orbital mechanism closes the \emph{same-spin} gap between the down-spin conduction and valence bands
[Fig.~\ref{fig:V_vs_Phi}(b)]. Because the Zeeman shifts cancel exactly
in this channel (SM Sec.~S-V.1), this closure is driven entirely by the
orbital Peierls phase. The directional vector potential renders the three satellite valence
maxima inequivalent, forcing an asymmetric distortion of the energy contours.

Because the three satellite valence maxima become inequivalent, the
same-spin gap is determined solely by the highest of these maxima and
the central conduction minimum. Since these extrema occur at distinctly
different momenta, the gap is indirect:
\begin{equation}
  \Delta_{ss}(B_\parallel) = \min_{\mathbf{k}} E_c^\downarrow(\mathbf{k}, B_\parallel) - \max_{\mathbf{k}} E_v^\downarrow(\mathbf{k}, B_\parallel).
  \label{eq:GapSS}
\end{equation}
If the system were subjected only to the Zeeman effect, this same-spin 
gap would remain rigidly fixed, as the Zeeman coupling merely shifts the 
entire spin-$\downarrow$ manifold as a single unit. It is the orbital 
vector potential that actively distorts the band dispersions, pushing the 
conduction minimum and the highest valence maximum toward each other 
in energy.

The orbital critical field $B_c^{\rm orb}$ is the precise point at which 
this asymmetric distortion forces $\Delta_{ss}$ to zero. For fields exceeding 
$B_c^{\rm orb}$, the conduction minimum drops below the valence maximum, 
causing the bands to overlap in energy and form distinct same-spin electron 
and hole pockets. We define the critical crossing energy, $E_L$, as the 
 ``charge-neutrality level" for the same-spin (down) subsystem: this is the specific 
energy at which these newly formed pockets are perfectly compensated. Because 
the density of states near these shallow band edges is approximately symmetric, 
$E_L$ is closely approximated by the midpoint of the overlapping extrema:
\begin{equation}
  E_L(B_\parallel) \approx \frac{1}{2} \left[ \min_{\mathbf{k}} E_c^\downarrow(\mathbf{k}, B_\parallel) + \max_{\mathbf{k}} E_v^\downarrow(\mathbf{k}, B_\parallel) \right].
  \label{eq:EL_midpoint}
\end{equation}

Crucially, this entire orbital band-overlap occurs roughly $1\,\mathrm{meV}$ 
below the global charge-neutrality point. Because $E_L$ is a thermodynamic 
property fixed entirely by the applied magnetic and displacement fields, 
the emergence of these new same-spin pockets is completely invisible to standard 
transport at half-filling; the massive cross-spin bands at the global 
Fermi level ($E_F$) dominate the signal. 

\begin{figure*}[ht]
  \centering
  \includegraphics[width=\textwidth]{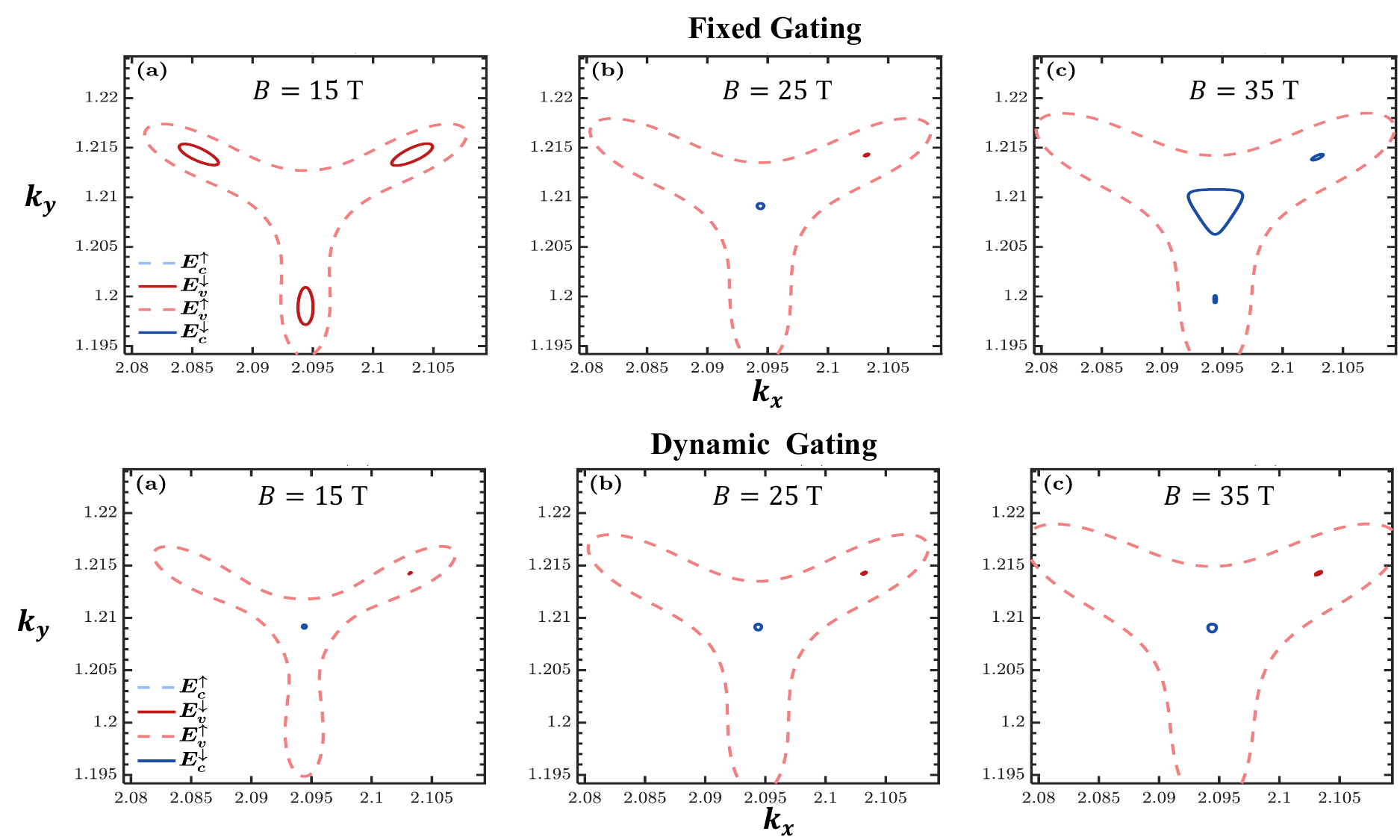}
 \caption{Orbital-driven $C_3$ symmetry breaking and emergence of same-spin Fermi pockets 
at $V=0.65\,\mathrm{meV}$.  Upper row: 2D energy contours at a constant energy $E_L(25\,\mathrm{T})$ for three different magnetic fields, $B=15$, $25$, and $35\,\mathrm{T}$. The same-spin Fermi pockets of electrons and holes are  visible only in the middle panel for $B \sim 25\,\mathrm{T}$.
Lower row: 2D energy contours evaluated  at a field-dependent energy allow the same-spin Fermi  pockets to remain visible over a broad range of magnetic fields. Continuously tuning the gate voltage to stay locked on $E_L$ reveals the true topological evolution of the same-spin pockets. The orientation of these emergent contours is highly sensitive to the azimuthal angle of $\mathbf{B}_\parallel$, rotating to preserve $C_3$ symmetry (see SM, Fig.~S2~\citep{SI}).}
\label{fig:FermiOrbital}
\end{figure*}

To observe the spin-resolved Fermi surface topology, the system must be explicitly gated. 
The experimentalist must tune the gate voltage to shift the chemical 
potential down until it coincides with the field-dependent value of $E_L$. 
Once the Fermi level is shifted to this specific energy, the $C_3$-broken same-spin Fermi 
contours emerge as contributors to transport. The appearance of these new band edges manifests 
as a distinct step-like onset in the same-spin DOS at $E_L$, as shown in 
Fig.~\ref{fig:V_vs_Phi}(b). These $C_3$-broken same-spin Fermi contours are the direct signature of
this emergent same-spin topology, as shown in
Fig.~\ref{fig:FermiOrbital}. The orbital transition involves
hybridization of the satellite Dirac cone wavefunctions and changes the
Berry curvature distribution of the filled
bands~\citep{PhysRevLett.99.236809,PhysRevLett.100.036804}. The
satellite cones carry Berry phase $\pi$~\citep{Novoselov2005,Zhang2005},
so the semimetallic pockets formed above $B_c^{\rm orb}$ are
topologically non-trivial, detectable as a half-integer offset in
Shubnikov-de Haas oscillations under a weak perpendicular
field~\citep{PhysRevLett.82.2147,Novoselov2005,Zhang2005}. The
sharp band-edge onset at $E_L$ produces the step-like feature around $-1.9$~meV in the same-spin DOS
shown in Fig.~\ref{fig:V_vs_Phi}(b).

\section{Conclusion \& Discussion}

In this work, we have demonstrated that incorporating trigonal warping ($\gamma_3$) and particle-hole asymmetry ($\gamma_4$) into the tight-binding description of AB-stacked bilayer graphene reveals a rich, dual-transition phase diagram under crossed electric and magnetic fields. Because the Zeeman and orbital degrees of freedom couple to the in-plane magnetic field at drastically different energy scales ($2\mu_B \gg \beta$), the system undergoes two separate topological transitions. 

The first transition occurs at low field ($B_c^Z \approx 1$\,T), where the Zeeman effect closes the cross-spin gap precisely at the ungated charge neutrality point ($E_F$). This causes an insulator-to-semimetal transition, yielding a spin-polarized semimetal. As the field is increased to $B_c^{\rm orb} \approx 10$--$25$\,T, the orbital vector potential continuously distorts the band structure until the same-spin gap closes. Crucially, this orbital-driven topological transition occurs at an energy $E_L$ located roughly $1$\,meV away from the ungated Fermi level. Consequently, the topologically nontrivial emergence of the asymmetric same-spin pockets remains hidden from standard half-filled transport and can only be accessed by explicitly gating the system to $E_L$. Our low-energy treatment \cite{SI} reveals an almost linear dependence of both critical fields ($B_c^Z$ and $B_c^{\rm orb}$) on the displacement field $V$, which we expect to saturate at higher fields once the magnetic-field-induced momentum shift exceeds the characteristic scale set by trigonal warping. 

Tuning the Fermi level to $E_L$ unveils a plethora of coexisting spin-energy contours. Extracting the signature of the emergent same-spin pockets against the immense spectral background of the already-established cross-spin valence contour poses an experimental challenge, but one that can be systematically addressed. In quantum oscillation (Shubnikov--de Haas) studies, the giant cross-spin contour will produce a high-frequency background, whereas the nascent same-spin pockets will contribute distinct, ultra-low-frequency oscillations, allowing clean isolation via Fourier analysis~\citep{PhysRevLett.104.066801}. Alternatively, scanning tunneling spectroscopy or gated capacitance measurements can directly resolve the sharp band-edge onset associated with the emergent same-spin pockets against the smoothly varying background density of states.

Beyond detection, the compensated, multi-pocket nature of these semimetallic phases provides a natural platform for exploring multi-component transport. The threefold-mini-pocket topology suggests the possibility of tuning carrier imbalances via suitable strain or twist, extending recent observations of two-fluid hydrodynamic conduction~\citep{Tan2022SciAdv, Shilov2024ACSNano, PhysRevB.97.121404} to a highly tunable, multi-valley, spin-selective regime. Furthermore, the small native gaps characterizing this system place it in a regime where excitonic instabilities have been predicted to occur~\citep{Scammell2023, PRB111_085433}. The field-driven generation of spin- and valley-resolved semimetallic pockets suggests that any competing excitonic states would inherit this rich internal structure, potentially realizing chiral or band-geometry-induced bound states~\citep{ValleyChiralExciton2026} as the Fermi surface evolves.
More immediately, the central role of skew-hopping in lowering $B_c^{\rm orb}$ suggests that other two-dimensional moir\'{e} and multilayer systems---where such couplings are inherently strong---may exhibit similarly accessible field-driven transitions. Our study provides a theoretical blueprint for engineering tunable, spin-textured band structures in van der Waals materials. Experimentally, gated bilayer-graphene devices can now be probed in pulsed or hybrid magnets up to $\sim30\,\mathrm{T}$ to directly observe both transitions via measurements of longitudinal conductivity, thermopower, and quantum oscillations.

\section*{Acknowledgments}
We thank Leonid Levitov for bringing to our attention
Ref.~\cite{Seiler2024}, which motivated this work. We also thank
Carsten Ullrich for a useful discussion on the role of the Zeeman
coupling, and Guang Bian for many useful discussions. A.C.\ was
supported by the U.S.\ Department of Energy, Office of Science, Office
of Basic Energy Sciences, Division of Materials Science and
Engineering, under Grant No.\ DE-SC0024294. G.V.\ was supported by the
Ministry of Education, Singapore, under its Research Centre of
Excellence award to the Institute for Functional Intelligent Materials
(I-FIM, Project No.\ EDUNC-33-18-279-V12).

\section*{Data Availability}
The code used to generate all numerical results and figures in this
work is publicly available at Ref.~\cite{code_repo}.

\clearpage
\onecolumngrid
\begin{center}
    \vspace{0.2cm}
    \textbf{\large Supplemental Material:\\Low-Field Metal-Insulator Transition in AB-Stacked Bilayer Graphene}
    \vspace{0.4cm}
\end{center}
\twocolumngrid

\setcounter{equation}{0}
\setcounter{figure}{0}
\setcounter{table}{0}
\setcounter{page}{1}
\setcounter{section}{0}
\renewcommand{\theequation}{S\arabic{equation}}
\renewcommand{\thefigure}{S\arabic{figure}}
\renewcommand{\thesection}{S-\Roman{section}}


\section{Rigorous Evaluation of the Interlayer Peierls Phase}

To evaluate the effect of an in-plane magnetic field on the tight-binding Hamiltonian, we calculate the Peierls phase accumulated during interlayer electron hopping. For an in-plane magnetic field applied along the $y$-direction, $\mathbf{B}=B_y\hat{y}$, we choose the Landau gauge such that the vector potential is $\mathbf{A}=B_y z\hat{x}$.

The magnetic phase modification to the hopping amplitude between two sites is given by the Peierls substitution line integral:
\begin{equation}
\Delta\phi = \frac{e}{\hbar}\int_{\mathbf{r}_i}^{\mathbf{r}_f} \mathbf{A}\cdot d\mathbf{l}
\end{equation}
We evaluate this for both the direct vertical hopping ($\gamma_1$) and the skew interlayer hoppings ($\gamma_3,\gamma_4$) between the two layers located at $z=\ell/2$ and $z=-\ell/2$.

\subsection{Vertical Interlayer Hopping ($\gamma_1$)}
The primary dimer coupling $\gamma_1$ connects an $A_1$ atom directly above a $B_2$ atom; the hop is strictly vertical. The differential path vector is $d\mathbf{l}=\hat{z}\,dz$. Because $\mathbf{A}$ lies entirely in the $x$-direction:
\begin{equation}
\Delta\phi_{\gamma_1}=\frac{e}{\hbar}\int_{\ell/2}^{-\ell/2} (B_yz\hat{x})\cdot(\hat{z}\,dz)=0.
\end{equation}

\subsection{Skew Interlayer Hopping ($\gamma_3$ and $\gamma_4$)}
For skew hoppings (e.g., $B_1\to B_2$ for $\gamma_3$), the path involves a vertical translation and an in-plane shift $\Delta\mathbf{r}_\parallel=\Delta x\hat{x}+\Delta y\hat{y}$. Parameterizing with $t\in[0,1]$:
\begin{align}
x(t) = x_0+t\Delta x,\quad y(t) = y_0+t\Delta y,\quad z(t) = \tfrac{\ell}{2}-t\ell.
\end{align}
The Peierls integral becomes
\begin{equation}
\Delta\phi_{\rm skew}=\frac{eB_y\Delta x}{\hbar} \int_0^1\Big(\frac{\ell}{2}-t\ell\Big)dt =\frac{eB_y\Delta x}{\hbar}\Big[\frac{\ell}{2}t-\frac{\ell}{2}t^2\Big]_0^1 = 0.
\end{equation}

\subsection*{Conclusion}
Because $A_x\propto z$, the vector potential is odd about $z=0$. The phase accumulated in the upper half of the bilayer exactly cancels that in the lower half. Consequently, $\Delta\phi_{\gamma_1}=\Delta\phi_{\gamma_3}=\Delta\phi_{\gamma_4}=0$. The gap-closing transition is driven entirely by the relative kinematic shift of the intralayer momenta via $A_x(z=\pm\ell/2)=\pm B_y\ell/2$.

\section{Two-Band Effective Hamiltonian under Crossed Fields-Excluding Zeeman}
\label{sec:C1}

To capture the detailed effect of trigonal warping we project the full $4\times4$ Hamiltonian onto the low-energy ($A_t$, $B_b$) subspace. The interlayer coupling $\gamma_1$ forms a dimer ($B_t\to A_b$), leaving the non-dimer sites ($A_t,B_b$) as the low-energy degrees of freedom. Rearranging into the block form $H=\bigl[\begin{smallmatrix}H_{11}&H_{12}\\H_{21}&H_{22}\end{smallmatrix}\bigr]$, the Schur-complement effective Hamiltonian is
\begin{equation}
\hat{H}_{1} = H_{11} - H_{12} H_{22}^{-1} H_{21}.
\end{equation}
%
However, before we proceed to do a discussion in low energy properties which are dominated by the band structure near the two valleys we need to expand the $f_k$-function near the valley $K$. Expanding $f(\mathbf{k})$ near the $K$ valley and retaining terms linear in $|\gamma_3|,|\gamma_4|,|V|,|\Delta|$ and quadratic in momentum, we obtain the two-band Hamiltonian~\eqref{eqn:heff22}. In the low-energy approximation the $f(k)$ reduces to 
\ber
f(k) &\approx-\frac{\sqrt{3}a}{2\hbar} (k_x-i k_y)+\frac{a^2}{8\hbar^2}\left(k_x+i k_y\right)^2 \\
&=-\frac{\sqrt{3}a}{2\hbar} \pi^\dagger++\frac{a^2}{8\hbar^2}\pi^2 
\eer
where we represent \(\pi= k_x+i k_y\) and \(\pi^{\dagger}= k_x-i k_y\). Also, we define \(v=\frac{\sqrt{3}a}{2\hbar}\gamma_0\),  \(v_3=\frac{\sqrt{3}a}{2\hbar}\gamma_3\) and \(v_4=\frac{\sqrt{3}a}{2\hbar}\gamma_4\). 
The in-plane field along $y$ enters as a real momentum shift: $f(\mathbf{k}\pm\phi\hat{x})=-\frac{\sqrt{3}a}{2\hbar}(\pi^\dagger\pm\phi)$, where $\phi=\pi\Phi=\pi Ba\ell/\Phi_0$. Including $V$:
\begin{widetext}
\begin{align}\label{eqn:heff22}
\hat{H}_{1} &= \begin{bmatrix}
h_{11} & h_{12} \\
h_{21} & h_{22}
\end{bmatrix}=
\begin{bmatrix}
V & v_3\pi-\tfrac{v_3^2}{6\gamma_3}(\pi^\dagger)^2 \\
v_3\pi^\dagger-\tfrac{v_3^2}{6\gamma_3}\pi^2 & -V
\end{bmatrix}\\
&-
\begin{bmatrix}
\tfrac{Vv^2}{\gamma_1^2}(\pi^\dagger+\phi)(\pi+\phi)-\tfrac{vv_4}{\gamma_1}[\pi(\pi^\dagger+\phi)+\mathrm{H.c.}]
 & \tfrac{v^2}{\gamma_1}[(\pi^\dagger)^2-\phi^2]-\tfrac{2V\gamma_0\gamma_4}{\gamma_1^2}(\pi^\dagger\phi) \\
\tfrac{v^2}{\gamma_1}[\pi^2-\phi^2]-\tfrac{2V\gamma_0\gamma_4}{\gamma_1^2}(\pi\phi)
 & -\tfrac{Vv^2}{\gamma_1^2}(\pi^\dagger-\phi)(\pi-\phi)-\tfrac{vv_4}{\gamma_1}[\pi(\pi^\dagger-\phi)+\mathrm{H.c.}]
\end{bmatrix}.
\end{align}
\end{widetext}
This deconstructs as $\hat{H}_1=h_0+h_w+h_a+h_V$ with 
\begin{align}
h_0&=-\frac{v^2}{\gamma_1^2}\begin{bmatrix}0&(\pi^\dagger)^2-\phi^2\\(\pi)^2-\phi^2&0\end{bmatrix},\\
h_w&=-v_3\begin{bmatrix}0&\pi-\tfrac{v_3}{6\gamma_3}(\pi^\dagger)^2\\\pi^\dagger-\tfrac{v_3}{6\gamma_3}\pi^2&0\end{bmatrix},\\
h_V&=\begin{bmatrix}V&0\\0&-V\end{bmatrix}
   -\frac{Vv^2}{\gamma_1^2}\begin{bmatrix}(\pi^\dagger+\phi)(\pi+\phi)&0\\0&-(\pi^\dagger-\phi)(\pi-\phi)\end{bmatrix}.
\end{align}

\section{Energy Dispersion and Gap Dynamics under Crossed Fields}

\subsection{Diagonal Elements: Renormalized Gap and Magnetic Tilt}
The diagonal elements arise from $h_V$. Transforming to polar coordinates $\pi=\xi pe^{i\xi\theta}$ and expanding the operator products in polar coordinates yields:
\begin{align}
h_{11} &= V - V \frac{v^2}{\gamma_1^2} \left( p^2 + \phi^2 + 2\phi \xi p \cos\theta \right) \\
h_{22} &= -V + V \frac{v^2}{\gamma_1^2} \left( p^2 + \phi^2 - 2\phi \xi p \cos\theta \right)
\end{align}
%
This expansion give a tilt term and a renormalized gap: $h_V = E_{\rm tilt}\mathbb{I}+\tilde{V}\sigma_z$, with
\begin{align}
E_{\rm tilt}&=-2V\frac{v^2}{\gamma_1^2}\phi\xi p\cos\theta,\quad
\tilde{V}=V-V\frac{v^2}{\gamma_1^2}(p^2+\phi^2).
\end{align}

\subsection{Off-Diagonal Elements: Warping and Flux}
The off-diagonal element is $h_{12}=\tfrac{\phi^2}{2m}-\tfrac{p^2}{2m}e^{-2i\xi\theta}+v_3\xi pe^{i\xi\theta}$ (where $m=\gamma_1/2v^2$), giving
\begin{align}
|h_{12}|^2 &= \underbrace{(v_3p)^2-\xi\tfrac{v_3p^3}{m}\cos3\theta
+\left(\tfrac{p^2}{2m}\right)^2}_{\text{Zero-field}}\\
&+\underbrace{\tfrac{\phi^4}{4m^2}-\tfrac{\phi^2p^2}{2m^2}\cos2\theta
+\xi\tfrac{\phi^2v_3p}{m}\cos\theta}_{\text{Flux corrections}}.
\end{align}

\subsection{Full Dispersion and the Indirect Gap}
The exact eigenvalues are
\begin{equation}
E_{c,v}(p,\theta)=E_{\rm tilt}\pm\sqrt{\tilde{V}^2+|h_{12}|^2}.
\end{equation}
The indirect sam-spin gap (relevant for the IM transition) is $E_{\rm indirect}=2\sqrt{\tilde{V}^2+|h_{12}|^2}-2|E_{\rm tilt,max}|$.

\subsection{Evaluation at the Mini-Pockets}
At the pocket radius $p_w\approx\gamma_1v_3/v^2$ and angles $\theta_w\in\{0,2\pi/3,4\pi/3\}$, the $\mathcal{O}(\phi^2)$ flux corrections destructively interfere: $\Delta\epsilon_1^2(\phi^2,p_w,\theta_w)=2\phi^2v_3^2[\cos\theta_w-\cos2\theta_w]=0$. The indirect gap then closes linearly,
\begin{equation}
\boxed{\Delta E_{\rm orb}=4V\!\left(\tfrac{v_3}{\gamma_1}\right)\phi},
\end{equation}
which directly gives Eq.~(6) of the main text upon converting $\phi$ to $B$. This leading-order result reproduces the correct parametric form $\beta\propto V$ but underestimates the numerical $\beta$ by $\sim18$, since $\gamma_3/\gamma_1\approx0.78$ is not small.

\section{Numerical Procedure for Phase Boundary Determination}
\label{sec:SM_numerics}

To map out the insulating and semimetallic regions in the $(V,\Phi)$ plane, we diagonalize the full $4\times4$ Hamiltonian $H(\mathbf{k};V,\Phi)$ [Eq.~(1) of the main text] on a fine $k$-grid centred at the $K$ point. For each $(V,\Phi)$ we extract the two middle eigenvalues $E_2\le E_3$.

We define the indirect (same spin) gap as
\begin{equation}
E_g(V,\Phi)=\min_\mathbf{k}E_3(\mathbf{k};V,\Phi)
            -\max_\mathbf{k}E_2(\mathbf{k};V,\Phi).
\label{eq:SM_gap_def}
\end{equation}
By convention $E_g>0$ is insulating and $E_g\le0$ signals a compensated semimetal. For fixed $\Phi$ we scan $V$ to locate $V_c(\Phi)$ at $E_g=0$. Because $V_c$ depends sensitively on $\gamma_3/\gamma_1$, a dense $k$-grid is required (see Sec.~S-VII for robustness against parameter variation).

\begin{figure}[ht]
  \centering
  \includegraphics[width=\linewidth]{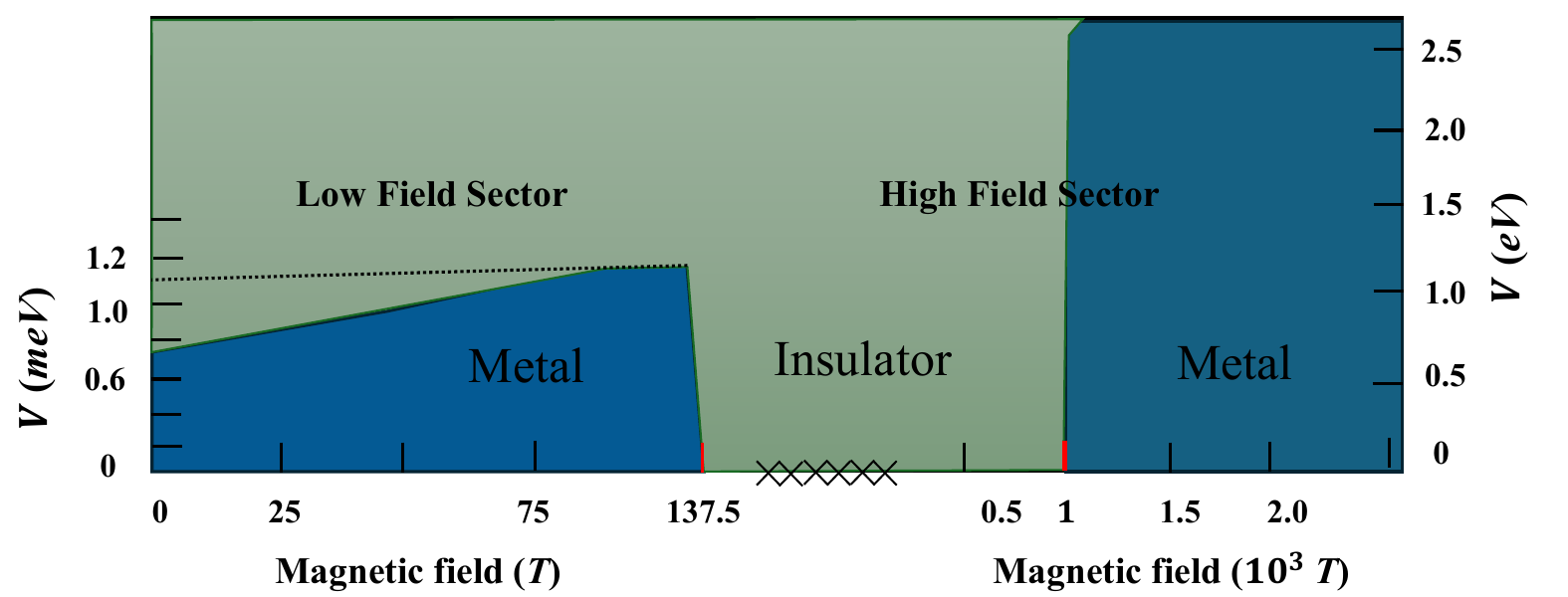}
  \caption{Full phase diagram in the $(V,\Phi)$ plane. The low-field sector exhibits an insulator-metal boundary driven by $V$ and $\Phi$; a crossover near $\sim137.5\,\mathrm{T}$ connects to the high-field unwarped regime of Ref.~\cite{PhysRevB.111.125130}. Note that the regions labeled `Metal' correspond to the sequential semimetallic phases detailed in the main text.}
  \label{fig:SM_full_phase}
\end{figure}

Figure~\ref{fig:SM_full_phase} shows the complete phase diagram spanning both low- and high-field regimes. The gap-vs-field curve at $V=0.7\,\mathrm{meV}$, together with the Zeeman cross-spin gap, is presented as Fig.~2(d) of the main text, since it directly supports the central claim of the paper.

\section{Effect of Zeeman Coupling}
\label{sec:SM_zeeman}

\begin{figure*}[ht]
  \centering
  \includegraphics[width=\linewidth]{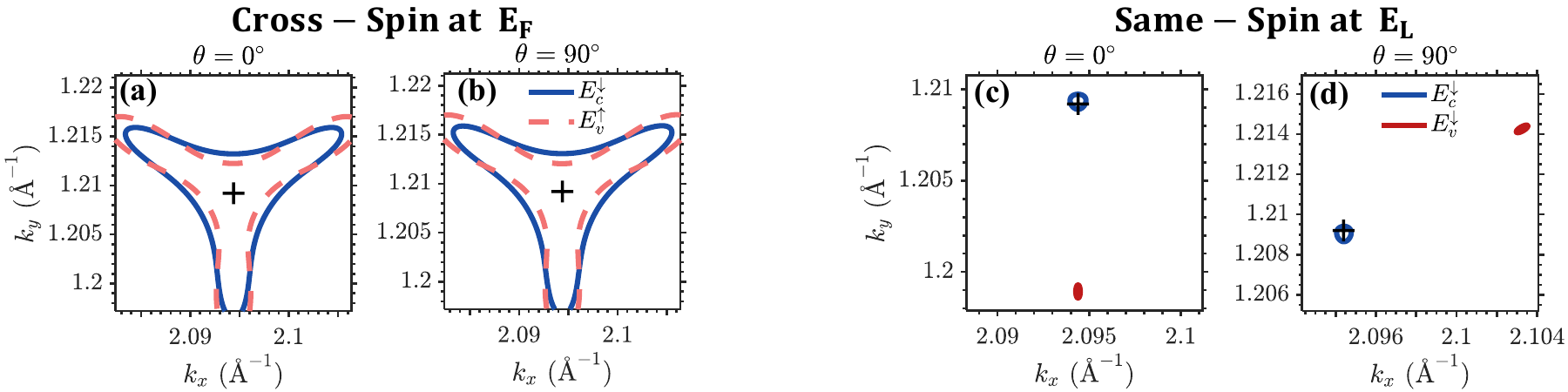}
  \caption{Spin-resolved Fermi contours at two in-plane field directions
$\theta=0^\circ$ and $\theta=90^\circ$, at $V=0.7\,\mathrm{meV}$.
\textbf{(a,b)}~Cross-spin bulk Fermi surface evaluated at the global
$E_F=0.43\,\mathrm{meV}$: the extended, three-lobed contour is
essentially unchanged between the two angles. \textbf{(c,d)}~Same-spin
pockets evaluated at the critical crossing energy $E_L$: the pocket
pair relocates from one satellite valley at $\theta=0^\circ$ to a
$C_3$-equivalent satellite valley at $\theta=90^\circ$, reflecting the
angle-dependent selection of the dominant valence maximum discussed in
the text --- a qualitatively different behavior from the isotropic
cross-spin contours in (a,b).}
  \label{fig:SM_angular}
\end{figure*}
As established in Eq.~(3) of the main text, the Zeeman coupling adds a uniform scalar shift $\sigma\mu_BB\,\mathbf{I}_{4\times4}$ to the Hamiltonian, where $\sigma=\pm1$ for spin-$\uparrow$/spin-$\downarrow$. Every eigenvalue shifts rigidly:
\begin{equation}
E_n^\sigma(\mathbf{k})=E_n(\mathbf{k})+\sigma\mu_BB,
\label{eq:SM_rigid_shift}
\end{equation}
where $E_n(\mathbf{k})$ are the spinless eigenvalues.

\subsection{Closing of the same-spin gap: exact cancellation of Zeeman contribution}
For a single spin species the indirect gap is
\begin{align}
E_g^\sigma
&=\min_\mathbf{k}E_3^\sigma(\mathbf{k})-\max_\mathbf{k}E_2^\sigma(\mathbf{k})
\nonumber\\
&=\min_\mathbf{k}[E_3(\mathbf{k})+\sigma\mu_BB]
 -\max_\mathbf{k}[E_2(\mathbf{k})+\sigma\mu_BB]
 =E_g(B),
\end{align}
where $E_g(B)$ is the spinless orbital gap [Eq.~(5) of the main text]. The Zeeman shift cancels exactly for either spin species. The Zeeman coupling therefore has no mathematical effect on the same-spin gap or on the orbital vector potential mechanism that drives its closure.

\subsection{Cross-spin gap and fine-tuning robustness}
The separation between the spin-$\downarrow$ conduction and spin-$\uparrow$ valence bands (the cross-spin gap) is
\begin{equation}
E_{c,\min}(\downarrow)-E_{v,\max}(\uparrow)=E_g(B)-2\mu_BB.
\label{eq:SM_cross}
\end{equation}
Setting this to zero and using $E_g(B)=\Delta(V)-\beta(V)B$ [Eq.~(5) of the main text] gives the threshold for the spin-polarized semimetal:
\begin{equation}
B_c^Z=\frac{\Delta(V)}{2\mu_B+\beta(V)}\approx\frac{\Delta(V)}{2\mu_B}.
\label{eq:SM_BcZ}
\end{equation}
This approximation holds because $\beta(V{=}0.7\,\mathrm{meV})\approx3.1\times10^{-3}\,\mathrm{meV/T}$ while $2\mu_B=0.1158\,\mathrm{meV/T}$. The orbital fraction governing this transition is thus:
\begin{equation}
r\equiv\frac{\beta}{2\mu_B+\beta}\approx 2.6\%.
\label{eq:SM_r}
\end{equation}
Numerically, at $V=0.7\,\mathrm{meV}$ ($\Delta(V)\approx 0.07\,\mathrm{meV}$), Eq.~\eqref{eq:SM_BcZ} gives $B_c^Z\approx0.6\,\mathrm{T}$, matching the exact numerical diagonalizations. 
Since $B_c^Z\propto\Delta(V)/(2\mu_B+\beta)$ and $B_c^{\rm orb}\propto\Delta(V)/\beta$, both depend linearly on $\Delta(V)$, giving:
\begin{equation}
\frac{dB_c^Z/dV}{dB_c^{\rm orb}/dV}=r\approx2.6\%
\;\Longrightarrow\;
\frac{\delta B_c^Z}{\delta B_c^{\rm orb}}\approx\frac{1}{38}.
\label{eq:SM_finetuning}
\end{equation}
For the same gate-voltage uncertainty $\delta V$, the absolute uncertainty in $B_c^Z$ is $\sim38\times$ smaller (in Tesla) than in $B_c^{\rm orb}$.

\subsection{The critical crossing energy $E_L$ offset}
The orbital transition at $B_c^{\rm orb}$ closes the same-spin gap at the critical crossing energy $E_L$. Using the symmetric midpoint approximation established in Eq.~(10) of the main text, with $E_3^\downarrow=E_3-\mu_BB$ and $E_2^\downarrow=E_2-\mu_BB$ (both Zeeman-shifted identically by $-\mu_BB$), one finds:
\begin{equation}
E_L \approx E_F - \mu_BB + \delta_{\gamma_4},
\label{eq:SM_EL2}
\end{equation}
where $\delta_{\gamma_4}\approx0.1\,\mathrm{meV}$ is the particle-hole asymmetry correction originating from $\gamma_4$. The energy separation between the bulk Fermi level and this emergent band-edge crossing is therefore:
\begin{equation}
\Delta E = E_F - E_L \approx \mu_BB.
\label{eq:SM_DeltaE}
\end{equation}
This gap grows linearly with the magnetic field. Consequently, the emergent same-spin pockets at $E_L$ are not located at the global bulk Fermi level $E_F$, but lie $\approx\mu_BB$ below it, necessitating targeted electrostatic gating for experimental observation.

\section{Angular Dependence of the Fermi Contours}
\label{sec:SM_angular}

The main text states that the critical field $B_c^{\rm orb}$ varies only
weakly with the direction $\theta$ of the applied in-plane field --- a
statement about the \emph{scalar threshold}, which by the exact $C_3$
symmetry of the lattice must in fact be periodic in $\theta$ with period
$120^\circ$. This periodicity does not, however, imply that the
Fermi-surface \emph{contours} themselves are angle-independent; the two
spin channels behave very differently under rotation of the field, for
a simple geometric reason.

The cross-spin Fermi surface at $E_F$ [Fig.~\ref{fig:SM_angular}(a,b)]
is a single, connected, three-lobed contour that wraps around all three
satellite positions simultaneously. Because this contour by construction
samples all three $C_3$-related directions at once, rotating the field
from $\theta=0^\circ$ to $\theta=90^\circ$ leaves its shape essentially
unchanged.

The same-spin pockets at $E_L$ [Fig.~\ref{fig:SM_angular}(c,d)] are, by
contrast, small and pointlike, sitting at whichever single satellite
valence maximum is currently highest in energy. The three satellite
maxima are degenerate only at $\phi=0$; a finite in-plane field tilts
them via the $E_{\rm tilt}\propto\cos(\theta-\theta_w)$ term of
Sec.~\ref{sec:C1}, and the maximum closest in angle to $\hat{\theta}$
becomes the dominant one. Since the three satellites are separated by
$120^\circ$, each dominates over a $60^\circ$-wide window of field
direction. Rotating the field from $\theta=0^\circ$ to $\theta=90^\circ$
is therefore sufficient to cross out of one satellite's window and into
a neighboring one: rather than rotating smoothly, the same-spin pocket
pair relocates discretely to the adjacent $C_3$-equivalent satellite, as
seen directly in Fig.~\ref{fig:SM_angular}(c,d).

This distinction --- a near-isotropic scalar threshold combined with a
strongly angle-selective pocket \emph{identity} --- is itself a
consequence of the differing topology of the two Fermi surfaces: an
extended, multiply-connected contour cannot easily discriminate between
directions, while an isolated point-like pocket is maximally sensitive
to exactly which extremum currently wins. In practice, this offers a
field-orientation handle for selecting which of the three momentum-space
valleys hosts the emergent same-spin carriers, independent of the field
magnitude required to access them.

\section{Parameter Robustness}
\label{sec:SM_robustness}

Because the critical field depends sensitively on $\gamma_3/\gamma_1$ (Sec.~S-IV), we assess the robustness of our central conclusions against (i) independent $\pm20\%$ variation of $\gamma_3$ and $\gamma_4$ about their literature values, and (ii) substitution of complete parameter sets from different sources.

\begin{figure}[ht]
  \centering
  \includegraphics[width=\linewidth]{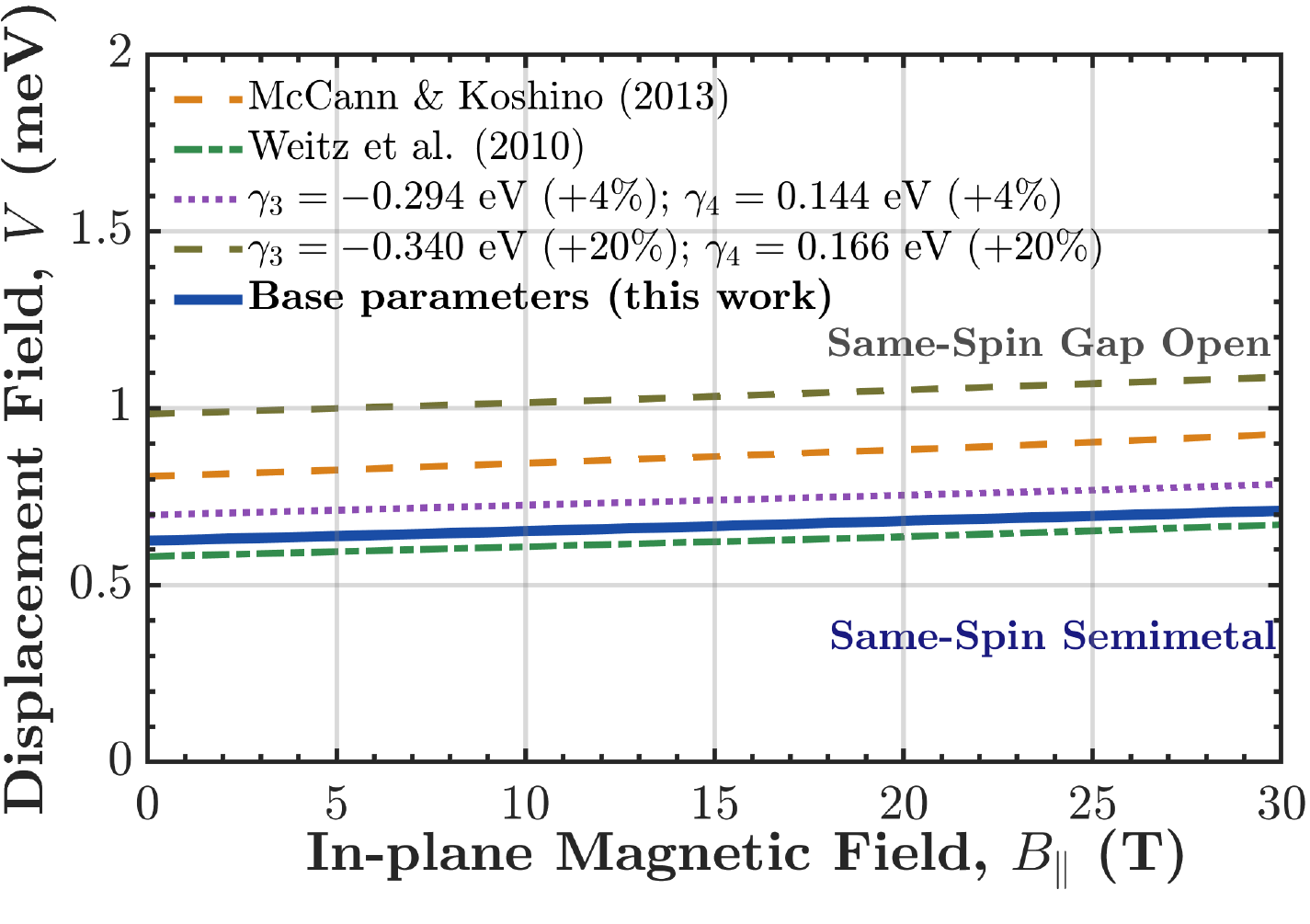}
  \caption{Orbital phase boundaries in the $V$--$B_\parallel$ plane evaluating parameter robustness. Each curve represents the critical boundary where the same-spin gap closes for a given set of skew hoppings. For any specific parameter set, the region above its curve corresponds to an open same-spin gap, while the region below corresponds to the same-spin semimetallic phase. To extract the critical orbital magnetic field $B_c^{\rm orb}$ for a chosen displacement field $V$, one traces horizontally from the $y$-axis to intersect the respective curve. Conversely, tracing vertically from a fixed $B_\parallel$ yields the threshold displacement field $V_c^{\rm orb}$ required to open the same-spin gap. The Zeeman boundary $V_c^Z$ is omitted here for visual clarity, as its sensitivity to $\gamma_3$ and $\gamma_4$ is negligible (entering only through $r\approx2.6\%$ [Eq.~\eqref{eq:SM_r}]).}
  \label{fig:SM_robustness}
\end{figure}

A subtlety arises when comparing $B_c^{\rm orb}$ across parameter sets at a single fixed displacement field. Because $\gamma_3$ and $\gamma_4$ both shift $V_c$ itself, evaluating $B_c^{\rm orb}$ at one fixed $V$ can place different parameter sets at very different distances above their respective thresholds, producing large apparent variation that does not reflect a change in the underlying mechanism. 

Three illustrative examples: at $\gamma_3=-0.2943\,\mathrm{eV}$ ($+4\%$), $V_c$ shifts to $0.677\,\mathrm{meV}$, leaving a gap $\Delta(V)=0.023\,\mathrm{meV}$ at $V=0.7\,\mathrm{meV}$ --- more than $3\times$ smaller than baseline ($0.07\,\mathrm{meV}$) --- which reduces $B_c^{\rm orb}$ to $\approx8.5\,\mathrm{T}$ at that fixed $V$. At $\gamma_4=+0.1546\,\mathrm{eV}$ ($+12\%$), $V_c=0.682\,\mathrm{meV}$ and $B_c^{\rm orb}\approx6.5\,\mathrm{T}$ at the same fixed $V$. For the McCann--Koshino (2013) set, $V_c=0.805\,\mathrm{meV}$ actually exceeds $0.7\,\mathrm{meV}$, so the fixed-$V$ comparison is undefined. 

In all three cases, evaluating $B_c^{\rm orb}$ at the \emph{same relative distance} $V=V_c+0.07\,\mathrm{meV}$ above the local threshold recovers $B_c^{\rm orb}\approx17$--$25\,\mathrm{T}$, consistent with baseline. The full phase-diagram overlay of Fig.~\ref{fig:SM_robustness} makes this explicit: each curve is shown across its full $V$ range, so the local $V_c$ and the $B_c^{\rm orb}(V)$ above it are directly visible. Across all parameter sets, $B_c^{\rm orb}$ remains in the $\mathcal{O}(10\,\mathrm{T})$ range for $V$ a few tenths of a meV above the local threshold, confirming the central conclusion of this work.

\bibliography{mybib}

\end{document}